**Progression in self-efficacy, interest, identity, sense of belonging, perceived recognition and effectiveness of peer interaction of physics majors and comparison with non-majors and Ph.D. students**

Kyle M. Whitcomb[1], Alexandru Maries[2], and Chandralekha Singh[1]

[1] *Department of Physics and Astronomy, University of Pittsburgh, Pittsburgh, PA, 15260*

[2] *Department of Physics, University of Cincinnati, Cincinnati, OH, 45221*

**ABSTRACT**

The importance of science beliefs such as self-efficacy, interest, identity, sense of belonging, perceived recognition and effectiveness of peer interaction in science education has been increasingly recognized in recent years. Self-efficacy, interest, and identity can be considered students' internal beliefs, and sense of belonging, perceived recognition and effectiveness of peer interaction relate to students' perception of the inclusiveness of the instructional environment. Prior studies in physics education regarding these beliefs have focused primarily on introductory physics courses. Here, we use five years of data from a validated survey administered to non-majors (in courses with physics majors) during their first year, physics majors throughout their undergraduate education, and first-year physics Ph.D. students at a large research university in the US. We find that physics majors in the first-year responded to the survey prompts more positively than their non-physics major peers who were in the same introductory courses, with the largest differences in perceived recognition, interest, and physics identity and somewhat smaller differences in self-efficacy, perception of peer interaction, and sense of belonging. Further, the average survey responses of physics majors for each belief remain largely constant over time from their first-year of the undergraduate curriculum through the last year and comparable to the Ph.D. students. This suggests that students are adjusting their interpretation of the survey items to match the current level of expertise expected of them in the course in which the survey was administered. One exception occurs in the second year, when peer interaction and sense of belonging reach a minimum. This may be the case because the second year is a particularly difficult time for students as they adjust to classes consisting primarily of physics majors. Moreover, physics identity dips to the lowest value in the fourth year when many students are contemplating continuing in physics beyond their undergraduate years or switching fields. We also find that, consistent with prior studies with introductory students, perceived recognition is the best predictor of physics identity for physics majors throughout their entire physics education, pointing to the importance of instructors making a concerted effort to recognize and affirm their students throughout their education.

# INTRODUCTION

In recent years, education researchers have been increasingly recognizing the importance of attending to students' beliefs in STEM domains to improve student outcomes (Broda et al., 2018, Marshman et al., 2018, Binning et al., 2020, Theobald et al., 2020) because regardless of student performance, their beliefs play a key role in their short and long-term plans for pursuing studies and careers in these domains. Physics beliefs commonly investigated in education research include self-efficacy, interest, physics identity, perceived recognition by others (such as instructors and teaching assistants), sense of belonging, and perception of effectiveness of peer interaction (henceforth, peer interaction). Self-efficacy, interest, and physics identity can be thought of as internal beliefs, because even though they can be influenced by the external environment, they relate to students' sense of who they are, their interest and their ability to succeed in a particular domain. Perceived recognition, sense of belonging and peer interaction provide insight into students' perception of the extent to which the learning environment is inclusive, which in turn can influence their internal beliefs and performance in a course. For example, prior research has shown that when instructional design explicitly addresses these motivational beliefs by creating an inclusive learning environment, student learning is improved and achievement gaps reduced (Binning et al., 2020), and conversely, when instruction does not address these motivational beliefs, achievement gaps can widen (Karim et al., 2018, Maries et al. 2020). Therefore, some researchers (Theobald et al., 2018) have called for additional research into what they have termed the "heads-and-hearts" hypothesis, suggesting that instruction is more likely to improve outcomes for all students when two key elements are incorporated: effective pedagogy and use of inclusive teaching practices that attend to students' beliefs.

# FRAMEWORK AND RESEARCH QUESTIONS

Much prior research (e.g., Hurtado and Carter, 1997, Marra, 2009, Yeager and Walton, 2011, Walton and Carr, 2012, Hazari et al, 2010, 2013a, 2013b, 2020, Aguilar, 2014, Trujillo and Tanner, 2014, Godwin et al., 2016, Nissen and Stemwell, 2016, Broda, 2018, Marshman, 2018, Kalender et al., 2019a, 2019b, Banchevsky, 2019, Lock et al., 2019, Binning et al., 2020) has investigated the motivational beliefs mentioned above (and defined below), both with regards to their interrelations, and the connection between these motivational beliefs and student learning. Based on this prior research, our framework conceptualizes that creating a classroom environment which attends to these beliefs, i.e., helps students feel like they belong in their major, have high self-efficacy, feel recognized as capable of excelling by their instructor, teaching assistants and peers and helps improve students' interest in learning the content is a stepping stone to fostering an "inclusive learning environment" (Theobald et al., 2020). It is therefore important to understand how the beliefs of students in a specific major evolve over time so that instructors have information about where students are at a given time, which can in turn help them devise strategies to support their learning. Since much prior STEM education research investigating these beliefs either focused on a student population at a given point in time (e.g., using survey data collected during one semester – Nissen and Shemwell, 2016, Hazari et al., 2010, Kalender et al., 2019a), or at two distinct points in time (e.g., first and second year – Marshman et al., 2018, first and last year – Hazari et al., 2020), or at multiple points in time but focusing on only one belief (e.g., self-efficacy – Marra et al., 2009), there is a need for research investigating the progression of multiple beliefs and also investigating the interrelations between these beliefs for a population of students, e.g., physics majors, throughout their undergraduate career. This is because creating positive changes in these beliefs should be an explicit goal of instruction at all levels since they can impact student short and long-term outcomes. Thus, administering validated surveys to investigate students' physics self-efficacy, interest, identity, sense of belonging, perceived recognition and peer interaction is important in addition to content-based assessments to get a holistic picture of the effectiveness of instruction.

Below, we define the beliefs we focus on because they have been found to be central in student outcomes and summarize prior research about their impact on outcomes:

*Self-efficacy*, is defined as belief in one's ability to succeed in a domain (Bandura, 1991). Many prior studies related to self-efficacy have shown that it is linked to many aspects of education, both in general and in physics specifically, including academic achievement, persistence in the major, and career choices after graduation (Marshman et al., 2018 Broda, 2018, Hazari et al., 2013b, Lock et al., 2019, Gonsalvez, 2018, Marra et al., 2009, Hazari et al., 2020, Nissen and Shemwell, 2016).

*Interest* is described by positive emotions people have about a domain, which enhance their curiosity and drive to learn about that domain. Prior research has linked interest to learning and has also shown that making the course content more relevant to students' lives can enhance their interest and learning (Hidi, 2006).

*Identity* in physics is described by the extent to which students see themselves as a physics person. Prior interviews suggest that students consider themselves to be a physics person when they perceive that they can excel in physics courses. Prior research has linked physics identity to self-efficacy, perceived recognition, and interest (Hazari et al., 2010, 2013a, Kalender et al., 2019a, 2019b), and has shown that physics identity can play a major role in the academic trajectory and future career choices of undergraduates (Hazari et al., 2013b, Lock et al., 2019, Danielsson, 2012, Gonsalves, 2018). Prior research on identity (Hazari et al., 2010, 2013a, Kalender et al., 2019a, 2019b) has proposed a framework which suggests that physics identity (also called "internal" identity) is informed by the following: (1) perceived recognition by instructors, teaching assistants, peers, and family (also called "external identity"); (2) competency belief which is closely tied to self-efficacy; and (3) interest in physics. Prior studies have found that perceived recognition is the strongest predictor of physics identity (Hazari et al., 2010, Kalender et al, 2019b), which implies that physics instructors can boost their students' identification as a physics person by making an effort to recognize them as capable of exceling in physics.

*Perceived recognition* relates to students' beliefs that other people see them as being capable in a particular domain. As discussed above, prior research suggests that physics perceived recognition is the strongest predictor of physics identity. For example, Kalender et al. found that physics perceived recognition explained students' self-efficacy, interest, and identity (Kalender et al. 2019b), which have all been shown to influence student success.

*Sense of belonging*, or *social belonging* has been described as a fundamental human need and feelings of exclusion can distract one from being fully present, such as not being able to take advantage of and completely engage in classroom learning opportunities. It appears on Maslow's original hierarchical set of five human needs (Maslow 1973) described as a need for people "to feel like they belong to and are accepted in a social group". Much has been written about social belonging since then (Yeager & Walton, 2011, Aguilar et al., 2014, Mahar et al., 2013, Banchevsky et al., 2019) and research has shown that students who feel they belong have more motivation and engagement along with greater academic success (Trujillo & Tanner, 2013, Hurtado & Carter, 1997, Binning et al., 2020, Walton & Carr, 2012).

*Peer interaction* relates to students' beliefs about the extent to which the interactions they have with their peers in a classroom are effective. In order to create a positive learning environment, it is important for the classroom climate to promote collaboration in a low-stakes environment, and a positive learning environment contributes to students' sense of belonging, which in turn contributes to students' improved performance.

The study presented here extends prior research by focusing on these beliefs for undergraduate physics majors throughout four years. The study also includes non-majors in first-year courses with majors and first-year physics Ph.D. students at the same university which is a different cohort of students and was included for benchmarking purposes. In particular, we investigate the trends in these beliefs from the first to the fourth year in the undergraduate major and compare them with first-year non-majors and Ph.D. students at the same institution. We collect survey data in all these years related to these beliefs and perform both descriptive and inferential statistics to answer the following research questions:

**RQ1.** What are the trends over time for physics majors' self-reported self-efficacy, interest, identity, sense of belonging, perceived recognition and peer interaction using a validated survey?

**RQ2.** How consistent are the relationships between physics identity and other beliefs for physics majors throughout their education?

**RQ3.** How consistent are the relationships between self-efficacy and other beliefs for physics majors throughout their education?

**RQ4.** How consistent are the relationships between sense of belonging and other beliefs for physics majors throughout their education?

**RQ5.** How do all of the findings for physics majors compare with non-majors in the same first-year courses and from those of Ph.D. students at the same university?

## METHODS

Using the Carnegie classification system, the US-based university at which this study was conducted is a public, high- research doctoral university, with balanced arts and sciences and professional schools, and a large, primarily residential undergraduate population that is full-time and reasonably selective with low transfer in from other schools.

**Survey**

All the survey items used in this study were adapted from other validated surveys (Activation Lab Tools, Glynn et al., 2011, Hazari et al., 2010, and PERTS, 2019). The items were re-validated at our institution for the specific student population investigated here (physics majors, non-majors, Ph.D. students) via individual interviews, factor analysis and Chronbach's alpha which are shown in Table 1 along with validation measures for the physics majors. Students responded to each item with the degree to which they agreed or disagreed with the statements on a Likert scale from 1–4 (1 being "strongly disagree" and 4 being "strongly agree"), with the exception of the first two items for "Interest" which had unique response options listed in the caption of Table 1 and the set of items relating to "Sense of Belonging" for which responses were on a 1–5 Likert scale.

The survey was administered during the first two weeks (pre-survey) and last two weeks of the year (post-survey) in lecture courses throughout the undergraduate physics curriculum (including to non-majors in first-year physics courses) as well as first-year Ph.D. courses. The survey was first administered in Fall 2015 and included questions on physics self-efficacy and interest, with additions to the survey in Fall 2017 when we expanded our framework to include physics identity and inclusiveness of the learning environment measured by sense of belonging, perceived recognition and peer interaction. The survey took approximately 10 minutes for students to complete. The university provided demographic information such as age, gender, and ethnicity or race information using an honest broker process by which the research team received the information without knowledge of the identities of the participants. These data were used to determine which students were physics majors by considering those students who had, at any point in their education, declared a major in physics.

**Survey validation**

In previous studies, the survey items were re-validated for first-year physics courses at our institution, in which the majority of students are non-physics majors (Kalender et al., 2019a, 2019b). Since the survey was already re-validated for first-year courses, we also re-validated it for majors and Ph.D. students and found very similar results to those from the previous studies. The interviews that were conducted confirmed that students at all levels were correctly interpreting the survey items. Next, we conducted an exploratory factor analysis (EFA) and found that the items group into the constructs labeled in Table 1 in a manner very similar to the previous studies. A follow-up confirmatory factor analysis (CFA) found that the specified factors based upon the EFA have "good" model fit (SRMR < 0.08, CFI > 0.90, TLI > 0.90) (Kline, 2011, Browne & Cudeck, 1992).

In Table 1 we report the standardized factor loadings ($\lambda$) from this CFA for each item, where $\lambda^2$ indicates the amount of variance explained in each item by the factors (Kline, 2011). We also report Cronbach's α for each factor with multiple items. Table 2 contains the matrix of Pearson correlation coefficients between the identified factors (Kline, 2011, Freedman, 2007). With the correlations all falling between 0.30 and 0.77, we find a good balance between these factors relating to one another without being so correlated that they are measuring the same construct (Cohen et al., 2003).

**Survey respondents**

We had 123 majors, and in the first year, a total of 3,591 non-majors answered the survey questions. The sample size varies from year to year and in some cases differs between constructs as items relating to some constructs were added in Fall 2017 as the framework was expanded. The non-majors (in first-year course with majors) were primarily engineering majors (70%), with the rest from mathematics and chemistry.

**Table 1.** The survey items used in this study along with the response options given to students. Each item is grouped into a set of items based upon the results of the factor analysis, and the associated motivational construct for each set of items is listed. For each construct with multiple items, Cronbach's α is provided as a measure of the coherence (internal consistency) of these items. For each item, the standardized factor loading $\lambda$ from a CFA model is listed. In the CFA, the physics identity item loaded on each of the other three constructs and so three $\lambda$ values are specified for each of the interest ($\lambda_{Int}$), perceived recognition ($\lambda_{Rec}$), and self-efficacy ($\lambda_{SE}$) factors. All items are on a Likert scale from 1 to 4 measuring the degree to which students agree with the statements (with 1 indicating strong disagreement and 4 indicating strong agreement), except the first two Interest items which have the following unique response options. Option set A = {Never, Once a week, Once a month, Every day}. Option set B = {Very boring, Boring, Interesting, Very interesting}. Items marked with a † are reverse-coded prior to analysis in order to align the scales of all items. * Responses are on a 1–5 instead of 1–4 scale Likert scale.

| Motivational construct | α | $\lambda$ | Question text |
|---|---|---|---|
| Interest | 0.82 | 0.52 | • I wonder about how physics works [option set A] |
| | | 0.69 | • In general, I find physics [option set B] |
| | | 0.73 | • I want to know everything I can about physics |
| | | 0.76 | • I am curious about recent discoveries in physics |
| | | 0.82 | • I want to know about the current research that physicists are doing |
| Perceived Recognition | 0.79 | 0.90 | • My family sees me as a physics person |
| | | 0.90 | • My friends see me as a physics person |
| | | 0.58 | • My physics instructor/TA sees me as a physics person |
| Self-Efficacy | 0.82 | 0.56 | • †Other people understand more than I do about what is going on in this physics course |
| | | 0.65 | • I am able to help my classmates with physics in the laboratory or in recitation |
| | | 0.68 | • †I get a sinking feeling when I think of trying to tackle tough physics problems |
| | | 0.70 | • I understand concepts I have studied in physics |
| | | 0.68 | • If I wanted to, I could be good at physics research |
| | | 0.78 | • If I study, I will do well on a physics test |
| | | 0.73 | • If I encounter a setback in a physics exam, I can overcome it |
| Identity | N/A | $\lambda_{Int} = 0.24$ $\lambda_{Rec} = 0.27$ $\lambda_{SE} = 0.46$ | • I see myself as a physics person |
| Sense of belonging | 0.87 | 0.83 | • *I feel like I belong in this physics class |
| | | 0.78 | • *†I feel like an outsider in this physics class |
| | | 0.84 | • *I feel comfortable in this physics class |
| | | 0.68 | • *I feel like I can be myself in this physics class |
| | | 0.74 | • *†Sometimes I worry that I do not belong in this physics class |
| Peer Interaction | 0.94 | | My experiences and interactions with other students in this class… |
| | | 0.84 | • Made me feel more relaxed about learning physics |
| | | 0.94 | • Increased my confidence in my ability to do physics |
| | | 0.95 | • Increased my confidence that I can succeed in physics |
| | | 0.91 | • Increased my confidence in my ability to handle difficult physics problems |

The physics majors were identified in the university data because at the studied university, students have to declare their major at some point before graduating. All students who did so were included in the group of physics majors. Survey responses from the physics majors were grouped into "years" corresponding to the typical timeline of physics courses for students at the studied university. For example, responses from students in introductory physics 1 (taken in the first-year) are categorized as "year 1" while responses from students in modern physics (taken in the second year alongside several required mathematics courses) are categorized as "year 2." A majority of required physics courses for the majors – classical mechanics (CM), electricity and magnetism (EM), thermal physics beyond the introductory level – are taken in the third and fourth years and although it is recommended that students take CM and EM courses before other upper-level courses, some students do not follow this recommendation. This may be due to the increased use of higher-order mathematical rigor (e.g., use of partial differential equations, integration involving several variables) that is unavoidable in effectively dealing with CM and EM theory. However, for the purposes of this study courses are grouped by the recommended curriculum so responses from students in CM and EM courses are categorized as "year 3" and responses from students in quantum mechanics and thermal physics are categorized as "year 4". We note that some undergraduates may be included in multiple courses in the data over different years. However, the undergraduate and Ph.D. student cohorts, though at the same university, are entirely separate (hence, the Ph.D. group is used only for benchmarking purposes).

**Table 2.** The Pearson correlation coefficients ($r$) between the factors for each motivational construct in the CFA. Entries above the diagonal are omitted since the matrix is symmetric.

|  | Interest | Perceived recognition | Self-efficacy | Sense of belonging |
|---|---|---|---|---|
| Interest | 1.00 |  |  |  |
| Perceived recognition | 0.55 | 1.00 |  |  |
| Self-efficacy | 0.55 | 0.56 | 1.00 |  |
| Sense of Belonging | 0.42 | 0.45 | 0.77 | 1.00 |

**Analysis**

To investigate trends over time and answer **RQ1**, we restricted the sample to matched pre- and post-responses and report sample size, mean, and standard error for these matched responses. To answer **RQ2**, **RQ3**, **and RQ4** we conducted multiple linear regression models (Montgomery et al., 2012) on only the post-responses from students in each year, as has been done in previous studies in identity framework (Hazari et al., 2010, 2013a, 2013b, Kalender et al., 2019a). Since pre-responses do not appear in the models, we did not restrict to the matched responses in these multiple linear regression models, instead using all available post-responses (although the results are similar using post-responses for matched data). Lastly, all of the differences discussed explicitly are statistically significant at the $p < 0.05$ level and all the assumptions required for carrying out a multiple regression analysis hold, namely linearity of relationships, normally distributed residuals, and homoscedasticity and multi-colinearity is not an issue. For example, we conducted Shapiro-Wilks tests for each model and found the W statistics to be close to 1 with all but one (0.92) being greater than 0.97 indicating that the data can reasonably be approximated as normal. To answer **RQ5**, we use statistical tests (mainly a $t$-test) to compare the first-year majors with non-majors (who were taking the same courses, and were separated based on major) and also compare majors with Ph.D. students on all the motivational beliefs.

**RESULTS**

To answer **RQ1** and investigate the trends for physics majors' self-efficacy, interest, identity, sense of belonging, perceived recognition and peer interaction over time, we calculated the mean and standard error of each construct for majors in each year and plotted them in Fig. 1. We first discuss the internal beliefs (self-efficacy, interest, physics identity) for majors. We observe that interest (Fig. 1b) and self-efficacy (Fig. 1a) remain fairly consistent, although self-efficacy shows a slight drop from year 1 pre (at 3.17) to year 1 post (3.00), but after that point it remains relatively consistent through year 4, with slight decreases from each pre survey to the following post survey followed by a slight increase in the next year's pre survey. The maximum self-efficacy occurs in year 1 pre (3.17) and the minimum in year 3 post (2.88). Physics identity (Fig. 1c), while still largely consistent over time, shows a slight decline from the maximum 3.63 in year 1 pre to the minimum 2.98 in year 4 post (i.e., the span of undergraduate years).

With regards to students' perception of the inclusiveness of the learning environment, perceived recognition is very stable across years 1 through 4 (ranging from a minimum of 3.23 in year 1 post to a maximum of 3.40 in year 4 pre). The other two constructs are presented slightly differently than the other constructs in Fig. 1. Peer interaction (Fig. 1f) is only included on the post surveys since the survey items pertain to students' experiences and interaction with their peers throughout the course. Responses to peer interaction range from a minimum of 2.73 in year 2 post to a maximum of 3.05 in year 1 post. Moreover, it appears that the minimum reported peer interaction in year 2 post is somewhat misaligned with the rest of the responses, which are more consistently near 3 points on average. Similarly, with regards to sense of belonging (Fig. 1e, note the scale being 1-5), there is a quick decline from the maximum of 4.09 in year 1 pre to the minimum of 3.46 in year 2 post. However, this returns to 3.96 in year 3 pre and remains relatively consistent from that point forward. Considering both peer interaction and sense of belonging, the data suggests that the second year is a somewhat more difficult time for the students' relationships with their fellow physics majors. Interestingly, Fig. 1 shows that the internal beliefs including self-efficacy, interest, and identity are not as susceptible to change during the second year.

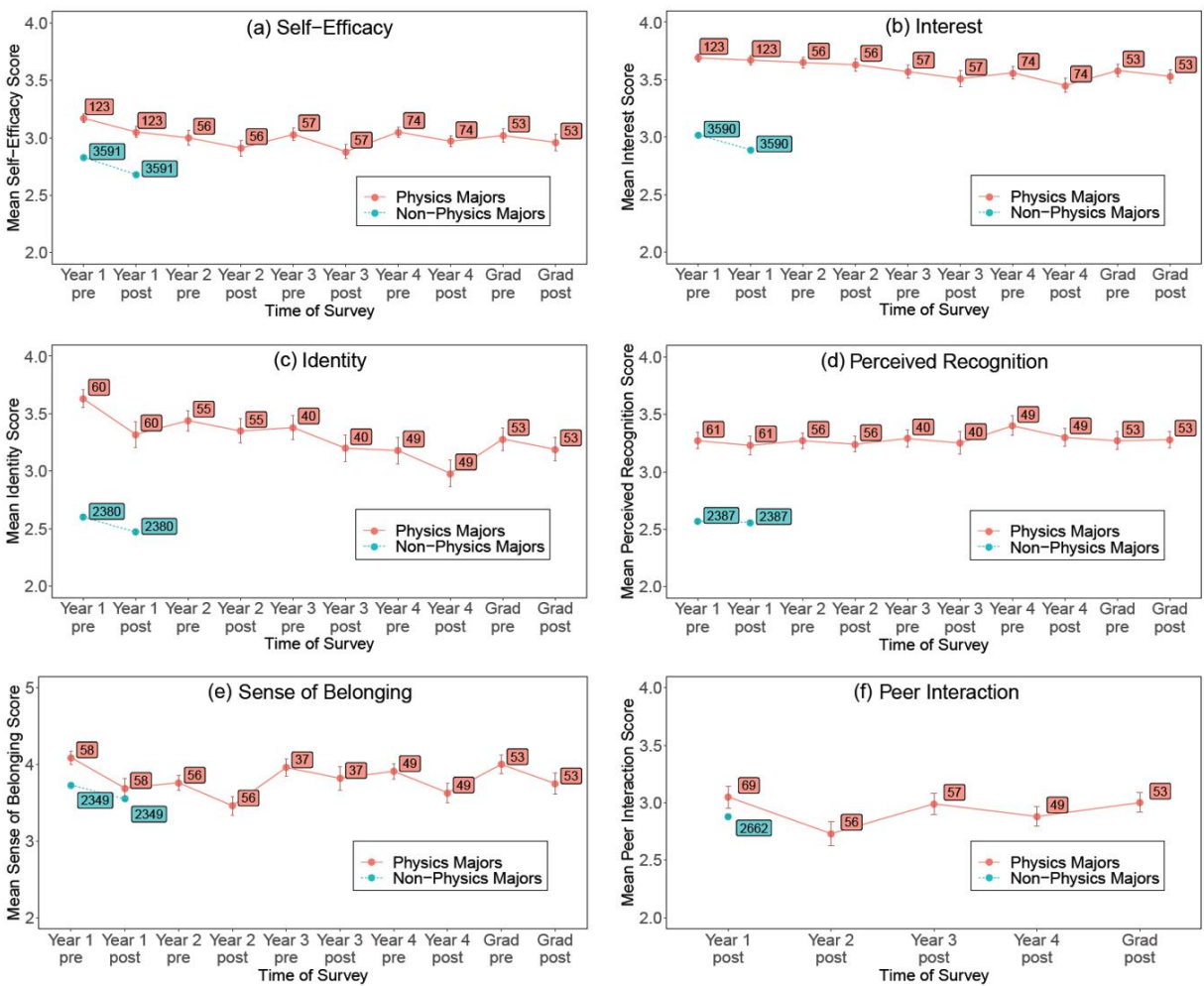

FIG. 1. The mean scores on (a) self-efficacy, (b) interest, (c) identity, (d) perceived recognition, (e) sense of belonging, and (f) peer interaction are plotted along with their standard error. Note that the responses for (a), (b), (c), (d) and (f) are on a Likert scale of 1–4 (with the plot restricted to 2–4 for visibility), and the responses for (e) are on a Likert scale from 1–5 (with the plot restricted to 2–5). The mean scores are plotted separately for the matched pre and post survey responses of physics majors in each of the four undergraduate years as well as first-year non-physics majors and first-year physics Ph.D. students ("Ph.D."). Since the questions pertaining to peer interaction were only given in the post survey, only post scores are plotted for those responses. The sample size is reported next to each point. Lines connecting points are drawn as guides to the eye.

Turning next to **RQ2** and investigating predictive relations, we use multiple regression models to investigate the extent to which physics identity is predicted by interest, perceived recognition, and self-efficacy in accordance with previous studies of physics identity (Hazari et al., 2010, 2013a, 2013b, Kalender et al., 2019a). Table 3 shows six such models, one for majors in each year of study, one for Ph.D. students, and one for non-majors (the data for Ph.D. students and non-majors will be discussed under **RQ5**). In every case, over 50% of the variance in physics identity is explained by the predictors (as indicated by $R^2$ in Table 3).

Focusing just on majors in year 1, we find a somewhat different model from what has previously been found for majors and non-majors (Hazari et al., 2010, 2013a, 2013b, Kalender et al., 2019a) wherein perceived recognition is the sole predictor of physics identity. We note, however, that there is also work to be done in improving student self-efficacy, as Fig. 1a shows that on average, a full point improvement is possible on the 1–4 Likert scale for the self-efficacy of physics majors. Progressing to year 2, for physics majors (Table 3), perceived recognition remains the top predictor of identity while self-efficacy becomes a statistically significant predictor. Following this, for majors in years 3 and 4, interest becomes a statistically significant predictor alongside perceived recognition and self-efficacy, with the standardized $β$ falling into a similar pattern as seen among non-majors in year 1.

**Table 3.** Summary of findings from multiple regression models predicting physics identity from interest, perceived recognition, and self-efficacy among physics majors in each year of study and non-physics majors in the first-year and Ph.D. students (Ph.D.). Each column shows the results from a regression using data for the specified year. Reported are the standardized regression coefficients ($β$) for each predictor along with the number of students ($N$) and fraction of variance explained in physics identity ($R^2$). The $p$-values are specified using asterisks, with non-significant predictors trimmed from the model and replaced by "–".

| Motivational Construct | Year_1 Non-majors | Standardized $β$ predicting physics identity | | | | |
|---|---|---|---|---|---|---|
| | | Year_1 | Year_2 | Year_3 | Year_4 | Ph.D. |
| Interest | 0.25*** | – | – | 0.32*** | 0.31*** | – |
| Recognition | 0.48*** | 0.74*** | 0.45*** | 0.42*** | 0.48*** | 0.62*** |
| Self-Efficacy | 0.21*** | – | 0.35*** | 0.20* | 0.22** | 0.34*** |
| $N$ | 2109 | 59 | 65 | 93 | 108 | 58 |
| $R^2$ | 0.64 | 0.54 | 0.52 | 0.63 | 0.63 | 0.74 |
| | | * $p < 0.05$ | ** $p < 0.01$ | *** $p < 0.001$ | | |

**Table 4.** Summary of findings from multiple regression models predicting physics self-efficacy from perceived recognition, peer interaction, and sense of belonging among physics majors in each year of study and non-physics majors in the first-year. The conventions and notations match those of Table 3.

| Motivational Construct | Year_1 Non-majors | Standardized $β$ predicting physics Self-Efficacy | | | | |
|---|---|---|---|---|---|---|
| | | Year_1 | Year_2 | Year_3 | Year_4 | Ph.D. |
| Recognition | 0.33*** | 0.50*** | 0.39*** | 0.36*** | 0.19* | 0.22* |
| Peer interaction | 0.16*** | – | 0.23* | – | – | – |
| Sense of belonging | 0.45*** | 0.45*** | 0.30* | 0.54*** | 0.64*** | 0.63*** |
| $N$ | 3013 | 76 | 65 | 70 | 69 | 57 |
| $R^2$ | 0.62 | 0.73 | 0.57 | 0.57 | 0.54 | 0.59 |
| | | * $p < 0.05$ | ** $p < 0.01$ | *** $p < 0.001$ | | |

**Table 5.** Summary of findings from multiple regression models predicting sense of belonging in physics from perceived recognition, self-efficacy, and peer interaction among physics majors in each year of study and non-physics majors in the first-year. The conventions and notations match those of Table 3.

| Motivational Construct | Year_1 Non-majors | Standardized $β$ predicting physics Sense of Belonging | | | | |
|---|---|---|---|---|---|---|
| | | Year_1 | Year_2 | Year_3 | Year_4 | Ph.D. |
| Recognition | 0.11*** | – | 0.29*** | – | – | – |
| Self-Efficacy | 0.52*** | 0.60*** | 0.26* | 0.53*** | 0.71*** | 0.60*** |
| Peer interaction | 0.22*** | 0.28*** | 0.43*** | 0.34*** | – | 0.27** |
| $N$ | 3013 | 76 | 65 | 70 | 69 | 57 |
| $R^2$ | 0.65 | 0.53 | 0.55 | 0.60 | 0.51 | 0.61 |
| | | * $p < 0.05$ | ** $p < 0.01$ | *** $p < 0.001$ | | |

To answer **RQ3** and **RQ4**, we tested the relationships between self-efficacy, peer interaction, and sense of belonging in Tables 4 and 5. Since these motivational constructs have complex inter-dependencies, we use these models to understand how these correlations evolve over time.

In Table 4, self-efficacy is primarily predicted by perceived recognition and sense of belonging, with peer interaction only being a significant predictor in year 2. As with Table 3, every model reported in Table 4 explains more than 50% of the variance in self-efficacy. In years 1 and 2, recognition and sense of belonging predict the self-efficacy of majors with roughly similar standardized $\beta$. However, among year 3 and 4, the best predictor of self-efficacy is sense of belonging, with the standardized regression coefficient $\beta$ of perceived recognition slowly decreasing over time, though always remaining statistically significant.

Finally, in Table 5, we investigate how sense of belonging depends on other beliefs. We find that in every model except for year 2 majors, the best predictor of sense of belonging is self-efficacy, with peer interaction being the next best predictor (except in year 4 when it is not statistically significant). The results for year 2 majors are once again an exception, as we saw when predicting self-efficacy in Table 4.

Turning now to **RQ5**, we note that for the comparison between majors and non-majors, the two sets of responses that appear in Fig. 1 for year 1 pre and year 1 post are from the same calculus-based physics courses, with those students who were identified as physics majors separated from the remaining non-physics majors. We find that on all internal beliefs, majors respond more positively on average than their non-major peers, though the difference between the two means is smaller in self-efficacy than in the other two constructs. In particular, for identity (Fig. 1c), the difference is 1.03 in pre and 0.85 in post and for interest (Fig. 1b), the difference is 0.67 in pre and 0.78 in post. However, majors and non-majors respond much more similarly in self-efficacy measures, with majors only 0.34 and 0.37 points higher on average in pre and post, respectively. The responses for peer interaction and sense of belonging are more similar between majors and non-majors than the items relating to perceived recognition (Fig. 1d) in which the difference is 0.70 in pre and 0.67 in post. In particular, physics majors only respond on average 0.17 points higher than non-majors on peer interaction in year 1 post, and on average 0.46 points higher in pre and 0.14 points higher in post on sense of belonging than non-majors.

The findings for non-majors in year 1 shown in Table 3 reflect prior research for introductory physics including both majors and non-majors (Hazari et al., 2010, 2013a, 2013b, Kalender et al., 2019a), and show that the primary predictor of physics identity is perceived recognition, while interest and self-efficacy are also statistically significant predictors (but with smaller regression coefficient $\beta$ than perceived recognition). For physics majors in year 1, only perceived recognition is a significant predictor of physics identity.

With regards to the comparison between majors and Ph.D. students, we note that the Ph.D. students have slightly higher physics identity and sense of belonging, but on the other measures (both with regards to beliefs and perception of the inclusiveness of the learning environment), the two groups show very similar averages. When looking at the predictive relationships shown in Table 3, we find interest once again drops out as a significant predictor of physics identity, leaving only perceived recognition and self-efficacy as the statistically significant predictors.

## DISCUSSION AND SUMMARY

In this study, which to our knowledge is the first such study investigating progression in beliefs, we find that, 1) perceived recognition is consistently the top predictor of physics identity across all years of the physics majors, and this is especially true in their first year, and 2) by and large, physics majors respond to the items pertaining to physics self-efficacy, interest, identity, sense of belonging, perceived recognition and peer interaction very consistently throughout their education (Fig. 1). The survey validation indicates that students were correctly interpreting these items and were responding to them in the context of the courses they were taking at the time, and their responses remained relatively constant. For example, first-year students responding to the prompt "if I study, I will do well on a physics test" responded to the item based on their belief about their performance on physics tests in their introductory physics course, while fourth-year students in a quantum mechanics course responded based on their belief about their performance on quantum mechanics tests.

It is interesting that these beliefs across the full physics curriculum are so consistent. However, there are some exceptions to this consistency. Physics identity drops noticeably from year 1 pre to year 1 post and year 4 pre to year 4 post. Both of these occur during extremely important moments in these students' careers: in year 1 these students are considering whether or not to continue their studies of physics, while in year 4 these students are deciding the next stage of their career after graduation. Less than 50% of students continue to graduate school, and only roughly 30% go on to physics graduate school. Thus, it makes sense that at the end of the fourth year, when we look at students' physics identity overall for the entire group, it would show a slight decline since it is reasonable to expect that some students who decide not to continue to physics graduate school (a decision typically made in year 4) would have lower physics identity. Another exception is a slight drop during year 2 in peer interaction and sense of belonging which may be due to the dramatic change in the physics class population between year 1 (which is largely dominated by engineering students) and year 2 (when the majority of students are physics majors). Another possible reason is that in the introductory year 1 courses, students attended recitation sections in addition to lecture, wherein they were among a smaller group of students from the class. These recitation sections are not present in physics courses beyond the first-year and so the year 2 courses may coincide with a transition period (which may even be marked by some students beginning to recognize the importance of establishing working groups among their physics major peers) while the students adjust to smaller class sizes in lecture instead of large lecture classes and smaller recitation sections.

Comparing majors with non-majors in their first year, we find consistently higher responses of majors. With regards to internal beliefs, this difference is larger in interest and identity (on the order of 0.6 to 1.0 on the 1-4 scale) than in self-efficacy (on the order of 0.3 on the 1-4 scale). With regards to perception of the inclusiveness of the learning environment, majors and non-majors show similar responses in sense of belonging and peer interaction, but there is a large discrepancy in perceived recognition (on the order of 0.7 on the 1-4 scale). One hypothesis for why this is the case is that even though physics is not their major, non-majors in these calculus-based physics courses are the majority, with most students being engineering majors, and so their sense of belonging and perception of positive peer interaction is bolstered by the presence of many peers from other common courses required by the engineering program. Additionally, we find that while for non-majors, interest, recognition, and self-efficacy are all significant predictors of physics identity, for majors, only recognition is a significant predictor. One hypothesis for this difference between majors and non-majors in year 1 is that majors have more homogeneous levels of interest and self-efficacy in physics while there is more variation in their perceived recognition.

Lastly, comparing physics Ph.D. students with majors, we find that the only differences in their physics identity and sense of belonging (higher for Ph.D. students), whereas on all other measures, they showed very similar averages. This may be in part due to PhD students comprising a different cohort than the undergraduates. Additionally, we find that while for majors in their last year, interest, recognition, and self-efficacy are all significant predictors of physics identity, for Ph.D. students, only recognition and self-efficacy are significant predictors. One hypothesis for this is that, as in the case with first-year undergraduates, first-year Ph.D. students have a more homogeneous level of interest in physics while their perceived recognition and self-efficacy vary.

## CONCLUSION AND IMPLICATIONS

Our findings strengthen and expand the findings of previous studies using the identity model (Hazari et al., 2010, 2013a, 2013b, Kalender et al., 2019a) which called for physics instructors in introductory courses to make a concerted effort to recognize their students as people who can excel in physics. In particular, we find that recognition is even more important for majors in the first-year than non-majors and Ph.D. students, and for undergraduates beyond the first-year, it is just as important as it is for non-majors in the first-year courses. We also found a decline in physics identity at crucial moments in students' undergraduate career, namely year 1 and year 4, and given that recognition is the top predictor of physics identity, this suggests that it may be useful to encourage majors by helping them recognize that they can excel in physics by working hard and providing appropriate support throughout the curriculum but especially during these transition periods. One promising approach instructors can use to help improve students' physics identity is via positive recognition; i.e., implementing measures to foster positive, encouraging interactions with all students from introductory to graduate levels. This is especially important for students who may have low self-efficacy and negative beliefs about their own ability to succeed. Research in introductory courses suggests that positive recognition from instructors can help create a learning environment in which all students can thrive (Hazari et al., 2010, Kalender et al., 2019b). Further, Tables 4 and 5 suggest that relationships among self-efficacy or sense of belonging and other constructs were consistent over time. Additionally, given the importance that prior research has placed on both self-efficacy and sense of belonging in supporting students' success and retention, efforts to foster positive recognition could be paired with promising interventions described in the literature focusing on self-efficacy and sense of belonging (Binning et al., 2020, Broda et al., 2018, Yeager & Walton, 2011, Aguilar et al., 2014). However, efforts to improve students' self-efficacy, as well as their sense of social belonging should be continued throughout the curriculum as needed to further solidify the impact of the interventions.

We note that this study is exploratory and more research is needed to confirm the analyses discussed here with regard to the differences found. Additionally, future studies after more data are collected could explore these relationships using other methods such as the Hierarchical Linear Model or Regression Tree.

## ACKNOWLEDGEMENTS

This research is supported by NSF Grant DUE-1524575 and is also discussed in K. M. Whitcomb's dissertation (Whitcomb, 2020).